\documentclass[11pt,a4paper]{emulateapj}
\usepackage{epsfig}
\usepackage{amsmath}
\usepackage{natbib}

\begin{document}

\title{A Measurement of the Black-Hole Mass in NGC 1097 using ALMA}

\author{K. Onishi\altaffilmark{1,2}, S. Iguchi\altaffilmark{1,2}, K. Sheth\altaffilmark{3}, and K. Kohno\altaffilmark{4,5}}
%\author{S. Iguchi\altaffilmark{1,2}}
%\author{K. Sheth\altaffilmark{3}}
%\and
%\author{K. Kohno\altaffilmark{4,5}}

\altaffiltext{1}{Department of Astronomical Science, SOKENDAI (The Graduate University of Advanced Studies), Mitaka, Tokyo 181-8588, Japan; kyoko.onishi@nao.ac.jp}
\altaffiltext{2}{National Astronomical Observatory of Japan, Mitaka, Tokyo 181-8588, Japan}
\altaffiltext{3}{National Radio Astronomy Observatory, 520 Edgemont Road, Charlottesville, VA 22903, USA}
\altaffiltext{4}{Institute of Astronomy, University of Tokyo, 2-21-1 Osawa, Mitaka, Tokyo 181-0015, Japan}
\altaffiltext{5}{Research Center for the Early Universe, University of Tokyo, 7-3-1 Hongo, Bunkyo, Tokyo 113-0033, Japan}

\vspace{2cm}
\begin{abstract}
We present an estimate of the mass of the supermassive black hole (SMBH) in the nearby type-1 Seyfert galaxy \object{NGC 1097} using Atacamma Large Millimeter/Submillimeter Array (ALMA) observations of dense gas kinematics.
Dense molecular gas dynamics are traced with ${\rm HCN} (J=1-0)$ and ${\rm HCO^{+}} (J=1-0)$ emission lines. 
Assuming a host galaxy inclination of $46^{\circ}$, we derive a SMBH mass, $M_{\rm BH}=1.40^{+0.27}_{-0.32} \times 10^{8}M_{\odot}$, and an I-band mass to light ratio to be $5.14^{+0.03}_{-0.04}$, using ${\rm HCN} (J=1-0)$. 
The estimated parameters are consistent between the two emission lines.
The measured SMBH mass is in good agreement with the SMBH mass and bulge velocity dispersion relationship. 
Our result showcases ALMA's potential for deriving accurate SMBH masses, especially for nearby late-type galaxies.  Larger samples and accurate SMBH masses will further elucidate the relationship between the black hole (BH) and host galaxy properties and constrain the coevolutionary growth of galaxies and BHs.
\end{abstract}

\section{Introduction}\label{sec:intro}
Recent observations suggest that supermassive black holes (SMBH) reside in the centers of most massive galaxies \citep[e.g.,][and references therein]{2013ARA&A..51..511K}. In the nearby universe a variety of host galaxy properties are known to be correlated with the central SMBH mass.  For instance there is a tight correlation between the SMBH mass and the bulge luminosity \citep[$M_{\rm BH}-L_{\rm bulge}$ relation, e.g.,][]{1995ARA&A..33..581K, 2013ApJ...764..184M, 2013ARA&A..51..511K}, the bulge mass \citep[$M_{\rm BH}-M_{\rm bulge}$ relation, e.g.,][]{1998AJ....115.2285M, 2003ApJ...589L..21M, 2012MNRAS.419.2497B, 2013ARA&A..51..511K}, and the central velocity dispersion \citep[$M_{\rm BH}-\sigma$ relation, e.g.,][]{2000ApJ...539L...9F, 2009ApJ...698..198G, 2013ApJ...764..184M, 2013ARA&A..51..511K}. These empirical correlations  suggest that black holes (BHs) may play a key role in the growth and evolution of galaxies. 

Recent studies, however, reveal that the correlation between the SMBH mass and bulge/host galaxy properties are more complex than originally thought.  For instance \citet{2013ApJ...764..184M} showed that the best-fit $M_{\rm BH}-\sigma$ relation differs between early- and late-type galaxies.  The difference results in the SMBH mass to be two times larger at a given velocity dispersion for the early-type $M_{\rm BH}-\sigma$ relation than for the late-type systems. In contrast, there does not seem to be a correlation between galaxies hosting psuedo-bulges  and their SMBH masses \citep{2013ARA&A..51..511K}. Much of the uncertainty comes from the scarce number of measurements of the SMBH masses especially in late-type galaxies, which is especially difficult with current methods for measuring the SMBH mass.

The most reliable way to estimate SMBH mass is to use dynamics/kinematics of gas or stars near the SMBH in a galaxy. So far, SMBH masses have been measured in the following ways: using proper motion of stars around the SMBH  \citep[e.g.,][]{2000MNRAS.317..348G, 2008ApJ...689.1044G}, using proper motion and dynamics of megamasers \citep[e.g.,][]{1995Natur.373..127M, 2011ApJ...727...20K}, stellar dynamics (usually only for elliptical or lenticular galaxies) \citep[e.g.,][]{1988ApJ...324..701D, 2011Natur.480..215M, 2014ApJ...791...37O}, and ionized gas dynamics \citep[e.g.,][]{1997ApJ...489..579M, 2008A&A...479..355D}. 

The method using the proper motion of stars around the central SMBH is the simplest of the four, but spatially resolving stars around a SMBH, besides the one in our Galaxy, is not currently possible. The SMBH mass measurements in galaxies using megamasers, which are rare and difficult to find, requires observations with very high angular resolution \citep[$\sim$ 0.3 milliarcsecond, for example, in ][]{1995Natur.373..127M}, accomplished with Very Long Baseline Interferometer. 
 Stellar dynamics have been used to measure the largest number of SMBH masses at this point. An orbit superposition model \citep{1979ApJ...232..236S} is fit to spectroscopic observations and the method is primarily restricted to elliptical and lenticular systems. The errors in this method can be amplified when a mass profile is asymmetric.
Ionized gas dynamics can be observed in larger samples (at least compared to the use of stellar or maser dynamics), by taking spectra of a galaxy using multiple slits or Integral Field Units (IFUs). The weak point of this method is that ionized gas is not necessarily settled into a pure rotating disk because it is more easily affected by non-circular motion from turbulence, shocks, radiation pressure, outflows, etc. Inferring the velocity field from slit observations is also problematic because such measurements may not always show non-circular motions which could be affecting the gas dynamics. Note that recent Integral Field Unit (IFU) observations can avoid this problem and are improving to be very useful to obtain velocity fields.

Deriving SMBH mass from molecular gas dynamics was not accomplished until \citet{2013Natur.494..328D} because millimeter-wavelength interferometers did not have sufficient angular resolution and sensitivity to map out the precise kinematics of molecular gas around a black hole. The SMBH mass in the central region of NGC 4526 was measured using the observed CO emission line from its circumnuclear, molecular-gas disk \citep{2013Natur.494..328D}. This method was also proposed by \citet{2013ASPC..476..275O}.
This technique is similar to the one used for ionized gas dynamics except that a molecular gas velocity field, which is primarily observed at millimeter/submillimeter wavelengths with interferometers, can  provide very high spectral resolution with full two-dimensional kinematics compared to optical or near-IR observations. Another major strength for molecular gas observations is that they can be utilized across the Hubble sequence from early- to late-type galaxies, as long as there is molecular gas in the galaxy centers. Thus molecular gas observations with new facilities like \textit{Atacamma Large Millimeter/submillimeter Array} (ALMA) offer a new and promising avenue for increasing SMBH measurements. A figure of merit for this method is summarized by \citet{2014MNRAS.443..911D}.

In this paper, we use the method initially employed by \citet{2013Natur.494..328D} and extend it to measure the SMBH mass using the ALMA data for \object{NGC 1097}(Project code = 2011.0.00108.S; PI = K. Kohno). The observations and data reduction are described in Section~\ref{sec:obs}. The SMBH mass measurement method is explained in Section~\ref{sec:model}. Section~\ref{sec:dis} contains discussions about the effect of dust extinction on the derived SMBH mass (Section~\ref{subsec:dis_dust}), how the inclination angle of NGC 1097 affects the result (Section~\ref{subsec:dis_warp}), the dependence of SMBH mass on the molecular gas species used to trace the dynamics (Section~\ref{subsec:dis_HCNHCO}). The conclusions are summarized in Section~\ref{sec:conc}.

\subsection{NGC 1097}\label{subsec:intro_n1097}
\object{NGC 1097} is a nearby Type-1 Seyfert galaxy at a distance of 14.5 Mpc \citep{1988ngc..book.....T} ($\sim70$~pc~arcsecond$^{-1}$). The position of the nucleus is determined by the peak position of the 6 cm continuum emission \citep{1987A&AS...70..517H}: $RA {\rm (J2000.0)} = 02^{\rm h} 46^{\rm m}18^{\rm s}.96$, $DEC {\rm (J2000.0)} = -30^{\circ}16'28''.9$. The peak position of the 860 $\mu$m continuum emission coincides with the 6 cm peak \citep{2013PASJ...65..100I}. Properties of \object{NGC 1097} are summarized in Table~\ref{table:n1097properties}.

The SMBH mass in \object{NGC 1097} is estimated to be $(1.2 \pm 0.2) \times 10^{8} M_{\odot}$ by \citet{2006ApJ...642..711L} using the empirical $M_{\rm BH}-\sigma$ relation from \citet{2002ApJ...574..740T} with an observed $\sigma=196 \pm 5$~km~s$^{-1}$. The uncertainty in this estimate is large, depending on the assumed $M_{\rm BH}-\sigma$ relation. The latest $M_{\rm BH}-\sigma$ relation [$\log_{10}(M_{\rm BH}/M_{\odot})=8.32+5.64\log_{10}(\sigma/200$ km~s$^{-1}$), \citet{2013ApJ...764..184M}] would yield SMBH mass of $(1.9 \pm 0.3) \times 10^{8}M_{\odot}$. Note that this relation is a fit to both late-type and early-type galaxies. When selecting only the late-type galaxies, the $M_{\rm BH}-\sigma$ relation becomes $\log_{10}(M_{\rm BH}/M_{\odot})=8.07+5.06\log_{10}(\sigma/200$~km~s$^{-1}$) \citep{2013ApJ...764..184M} and the estimated SMBH mass becomes $(1.1 \pm 0.3) \times 10^{8} M_{\odot}$. 

The enclosed mass in 40~pc radius has been studied by \citep{2013PASJ...65..100I} to be $2.8 \times 10^{8} M_{\odot}$, using the dynamics from ${\rm HCN} (J= 4-3)$ emission line. In contrast, \citet{2013ApJ...770L..27F} report a dynamical mass in $40 {\rm pc}$ radius as $8.0\times 10^{6}M_{\odot}$ from the same data of \citet{2013PASJ...65..100I}. The difference occurs because \citet{2013ApJ...770L..27F} assume a thin disk and extracts the non-circular motions of the gas while \citet{2013PASJ...65..100I} assume a simple Keplerian rotation. Note but the dynamical mass of \citet{2013PASJ...65..100I} includes all the mass within that radius, not showing the intrinsic SMBH mass. A more detailed study of \object{NGC 1097} is thus necessary to precisely measure the SMBH mass.

\begin{deluxetable}{lcc}
\tablecolumns{3}
\tablewidth{0pc}
\tablecaption{Properties of NGC 1097}
\tablehead{
\colhead{Parameter} &  \colhead{Value} & \colhead{Reference}}
\startdata
Morphology & SB(s)b & 1 \\
Nuclear activity & Type 1 Seyfert & 2 \\
Position of nucleus & \phn & 3 \\
\ \ \ $RA$(J2000.0) & $02^{\rm h}46^{\rm m}18^{\rm s}.96$ & \\
\ \ \ $DEC$(J2000.0) & $-30^{\circ}16'28''.9$ & \\
Systemic velocity (km~s$^{-1}$) & 1253\tablenotemark{a} & 4 \\
Position angle ($^{\circ}$) & 130 & 1, 4 \\
Inclination angle ($^{\circ}$) & 46$\pm$5 & 5 \\
Distance (Mpc) & 14.5 & 6 \\
Linear scale (pc~arcsec$^{-1}$) & 70 & 6 \\
I-band luminosity (mag) & 8.09 & 7 \\  
\enddata
%\vspace{-0.8cm}
\tablerefs{(1) \citet{1991rc3..book.....D}; (2) \citet{1993ApJ...410L..11S}; (3) \citet{1987A&AS...70..517H}; (4) this work; (5) \citet{1989ApJ...342...39O}; (6) \citet{1988ngc..book.....T}; (7) \citet{2007ApJS..172..599S} ;}
\tablenotetext{a}{Systemic velocity here is a heliocentric velocity determined with molecular lines. \citet{2004AJ....128...16K} shows the heliocentric velocity to be 1271~km~s$^{-1}$ determined with HIPASS observation. }
\label{table:n1097properties}
\end{deluxetable}

\section{Observations and Data Reduction}\label{sec:obs}
\begin{figure*}
\epsscale{1.0}
\plotone{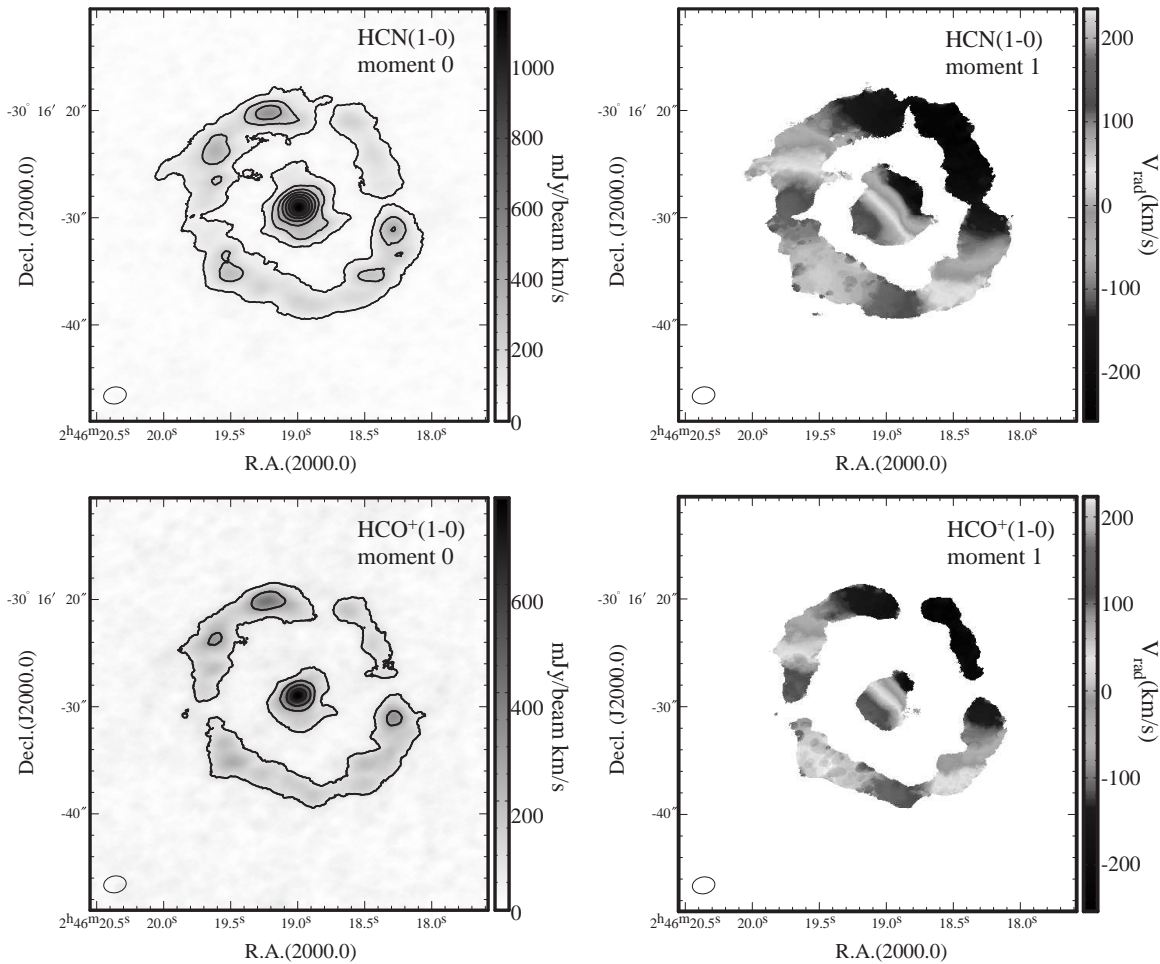}
\caption[Integrated Intensity Map and Intensity Weighted Velocity Map of ${\rm HCN} (J=1-0)$ and ${\rm HCO^{+}} (J=1-0)$]{\small \textit{(Left side)} Integrated intensity map (moment 0) for ${\rm HCN} (J=1-0)$ (upper panel, integrated for 88.1524-88.3467~GHz) and  ${\rm HCO^{+}} (J=1-0)$ (lower panel, 88.7139-88.9082~GHz). The rms noise level ($1\sigma$) in each integrated intensity map was 22~(mJy~beam$^{-1}$~km~s$^{-1}$) in ${\rm HCN} (J=1-0)$, and 26~(mJy~beam$^{-1}$~km~s$^{-1}$) in  ${\rm HCO^{+}} (J=1-0)$. The map is shown in gray scale with contour levels $3\sigma$ to $45\sigma$ in steps of $7\sigma$ for ${\rm HCN} (J=1-0)$ (upper panel), and $3\sigma$ to $24 \sigma$ in steps of $7\sigma$ for ${\rm HCO^{+}} (J=1-0)$ (lower panel). The synthesized beam size ($1''.60 \times 2''.20$ at $PA=-81.2^{\circ}$) is shown as the ellipse at the bottom left of each panel. 
\textit{(Right side)} The intensity weighted velocity map for ${\rm HCN} (J=1-0)$ (upper panel) and  ${\rm HCO^{+}} (J=1-0)$ (lower panel). Lower limit of the intensity is set to each map as $3\sigma$. The velocity resolution of each image is approximately 3.3~km~s$^{-1}$.}
\label{fig:HCNHCOmom01}
\end{figure*}

\object{NGC 1097} was observed with the band 3 receiver on ALMA using the two sideband dual-polarization setup as a cycle 0 observation (Project code = 2011.0.0108.S; PI = K. Kohno). The observations were conducted on 2012 Jul 29 and 2012 Oct 19 with an hour angle from $-4$ to $2$ and a total on-source time of 105.24 minutes. The antennas were in  the Cycle 0 extended configuration (400m baselines) which resulted in a synthesized beam of $1''.60 \times 2''.20$ at a position angle $-81.2^{\circ}$ ($\sim$ 112 pc $\times$ 154 pc). 
The receivers were tuned to cover the frequency range from 87.275~GHz to 100.917~GHz with two spectral windows each in the upper sideband (USB) and the lower sideband (LSB). Each spectral window had a bandwidth of 1.875~GHz with 3840 channels. The frequency resolution was 0.488~MHz per channel. Observational parameters are summarized in Table~\ref{table:obsproperties}. The field of view (full width half maximum of the primary beam) at these frequencies was $69''$. 
The data were reduced and imaged using CASA (Common Astronomy Software Applications) 4.0 with a robustness parameter of 0.5. We binned the data by 2 channels to improve the signal to noise ratio and our final resolution is 0.976~MHz or $\sim 3$~km~s$^{-1}$. Molecular gas emission is detected over 560~km~s$^{-1}$ (${\rm HCN} (J=1-0)$ emission is seen from  88.1524-88.3467~GHz and ${\rm HCO^{+}} (J=1-0)$ from 88.7139-88.9082~GHz). The integrated intensity moment zero and intensity weighted velocity maps were made using Karma \citep{1996ASPC..101...80G}. These are shown in Figure~\ref{fig:HCNHCOmom01}. The noise in the integrated intensity maps is 22~(mJy~beam$^{-1}$~km~s$^{-1}$) in the ${\rm HCN} (J=1-0)$ and 26~(mJy~beam$^{-1}$~km~s$^{-1}$) in the ${\rm HCO^{+}} (J=1-0)$ maps respectively. The peak flux is detected at $52 \sigma$ in the ${\rm HCN} (J=1-0)$ map  and at $29 \sigma$ in the ${\rm HCO^{+}} (J=1-0)$  map respectively. The data clearly show the rotation dominated kinematics of the molecular gas around the SMBH.

\begin{deluxetable}{lcc}
\tablecolumns{3}
\tablewidth{0pc}
\tablecaption{ALMA observation parameters}
\tablehead{
\colhead{Parameter} & \colhead{} & \colhead{}}
\startdata
Date & \multicolumn{2}{c}{2012 Jul 29, Oct 19} \\
On-source time & \multicolumn{2}{c}{105.24 minutes} \\
Configuration & \multicolumn{2}{c}{extended (Cycle 0)} \\
Phase center: & \multicolumn{2}{c}{\phn} \\
\ \ \ $RA$(J2000.0) & \multicolumn{2}{c}{$02^{\rm h}46^{\rm m}19^{\rm s}.06$} \\
\ \ \ $DEC$(J2000.0) & \multicolumn{2}{c}{$-30^{\circ}16'29''.7$} \\
Primary beams & \multicolumn{2}{c}{$69''$} \\
\phn & LSB & USB \\
\cline{1-3}\\
Frequency coverage (GHz) & 85.400-89.104 & 97.271-100.917 \\
Velocity resolution (km~s$^{-1}$) & 1.7 & 1.5 \\
Central frequency of \\each spectral window (GHz) & 86.338, 88.166 & 98.209, 99.979 \\
\enddata
\label{table:obsproperties}
\end{deluxetable}

\section{Supermassive Black-hole mass Estimation}\label{sec:model}
The measurement procedure for the SMBH mass and its result are described in this section. We model a mass distribution of the galaxy with multiple Gaussians to express the combination of stellar mass profile and the SMBH mass. The gravitational potential is derived by following the equations described in \citet{2002ApJ...578..787C}, which uses a Multi Gaussian Expansion (MGE) method. Circular velocity is calculated from the gravitational potential field by using MGE\_circular\_velocity code, which is in the JAM modelling package\footnote{http://purl.org/cappellari/idl} of \citet{2008MNRAS.390...71C}. The SMBH mass is estimated from the total mass profile, which gives a velocity field best matched to the observed result. When comparing the derived velocity profile with the observational results, we use the KINematic Molecular Simulation \citep[KinMS,][]{2013Natur.494..328D} in order to consider disk properties (e.g., disk thickness, position angle, and inclination) and the observational effect of beam-smearing. We show the details of each procedure in the following sections. 

\subsection{The Mass Model} \label{subsec:model_lumdis}
\begin{figure}
\epsscale{1.0}
\plotone{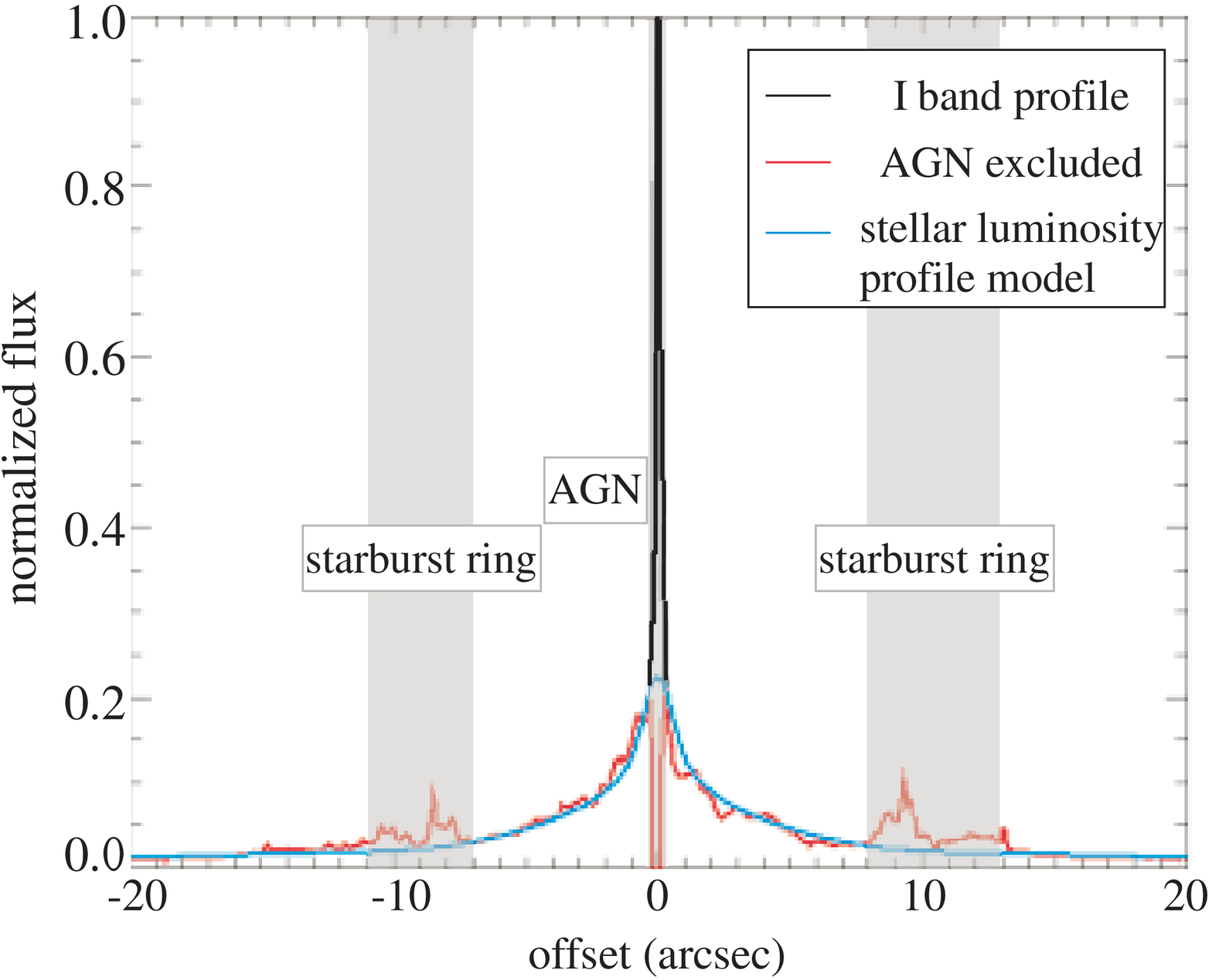}
\caption[\textit{HST} I-Band Luminosity Profile and Its Model]{\small The observed I-band luminosity distribution of \textit{HST} F814W (black line), the distribution with the AGN subtracted (red line), and the modeled stellar luminosity profile (blue line). We run the least-square value calculation at all radii excluding the colored regions to obtain the model profile. The starburst ring dominated regions (from $-11''$ to $-7''$ and from $8''$ to $13''$) and the central area where seems to have the luminosity dominated with the AGN (from $-0''.5$ to $0''.5$) are colored with gray. }
\label{fig:lumprof_Ibandmodel}
\end{figure}

We express the mass distribution as a summation of the SMBH mass and the stellar mass profile, expressed as the stellar luminosity distribution multiplied by a constant I-band mass-to-light (M/L) ratio.
Assuming that the galaxy is axisymmetric, the stellar luminosity distribution along the galaxy major axis is modeled as a superposition of Gaussian components, using the idea of Multi-Gaussian Expansion \citep[][]{1994A&A...285..723E}.
The major axis defined here is at a position angle of $130^{\circ}$ from \citet{1991rc3..book.....D}. 
We use an I-band image observed with the Advanced Camera for Surveys Wide Field Channel F814W filter on \textit{Hubble Space Telescope} (\textit{HST}) to obtain the luminosity distribution at the major axis (black line in Figure~\ref{fig:lumprof_Ibandmodel}).
We subtract the contribution from the Active Galactic Nucleus (AGN) and flatten the starburst dominated region (both are shaded in Figure~\ref{fig:lumprof_Ibandmodel}) in order to estimate the underlying stellar luminosity profile.
The AGN profile is calculated by the convolution of a delta function with the Point Spread Function (PSF), measured to be $0''.2$ (full width half maximum, FWHM) from five unresolved stars in the same image. The PSF is also checked by using ``Tiny~Tim" package (version~6.3) developed by \citet{2011SPIE.8127E..0JK}.
The luminosity value of the delta function is determined to obtain the residual luminosity distribution larger than 0 at any radius. Note that the AGN subtraction does not gravely affect the SMBH mass estimation. Even the peak before subtraction gives a mass of less than $6.00\times10^{6}M_{\odot}$ for a M/L ratio 5.14, and it is at least one digit smaller than the SMBH mass we put in the model. 
The starburst ring region is determined to be in the range of radii from $-11''$ to $-7''$ and from $8''$ to $13''$ along the major axis.
The ring includes younger stars than inside or outside of the ring, and is bluer in color. We model the stellar luminosity profile by calculating the least-square value with the luminosity distribution without the ring and the AGN. The model (blue line in Figure~\ref{fig:lumprof_Ibandmodel}), therefore, consists mainly of old stars, but does not include younger stars on the ring.
See also Table~\ref{table:MGE} for MGE parameters we give for the data. 
The mass profile is simply modeled by multiplying the constant I-band M/L ratio to the modeled stellar luminosity profile and adding the assumed SMBH point mass in the center. Note again that we model older stars, by which means we are assuming that the radial difference of the M/L ratio is negligible.

\begin{deluxetable}{lccc}
\tablecolumns{4}
\tablewidth{0pc}
\tablecaption{MGE Parameters for the Stellar Luminosity Obtained from WFC F814W Surface Brightness of NGC 1097}
\tablehead{
\colhead{$j$} &  \colhead{$I_{j}$($L_{\odot, I}/$pc$^{-2}$)} & \colhead{$\sigma_{j}$(arcsec)} & \colhead{$q_{j}$}}% 
\startdata
1.......... & 7451.23 & 0.500 & 0.900 \\
2.......... & 5122.72 & 3.50 & 0.700 \\
3.......... & 3725.62 & 1.25 & 0.900 \\
4.......... & 1862.81 & 16.0 & 0.700 \\
\enddata
\tablenotetext{a}{See text for how we deal with the AGN and the starburst region.}
\label{table:MGE}
\end{deluxetable}

\subsection{Velocity Field Calculation}\label{subsec:model_velcalc}
With a given mass profile from Section~\ref{subsec:model_lumdis}, we calculate the velocity field from a gravitational potential field derived with equations from \citet{2002ApJ...578..787C}. We use MGE\_circular\_velocity code, which is in the JAM modelling package of \citet{2008MNRAS.390...71C}.
The inclination angle is set to be $46^{\circ}$ \citep{1989ApJ...342...39O}. See Section~\ref{subsec:dis_warp} for more details regarding the inclination angle.

\subsection{Modelling a Position-Velocity Diagram (PVD)}\label{subsec:model_PVD}
We calculate a position-velocity diagram (PVD) model along the galaxy major axis by assuming that the observed molecular gas follows the velocity field obtained by our calculation.
The observational effects are taken into account by utilizing KinMS \citep[][]{2013Natur.494..328D}.
We convolve the model cloud distribution with our beamsize ($1''.60 \times 2''.20$) to express the beam-smearing, and assume the molecular gas disk to be an axisymmetric thin disk.
The position angle of the galaxy major axis is set to be $130^{\circ}$ \citep{1991rc3..book.....D}, which is consistent with the kinematical position angle estimated from our observational data. The position angle is also consistent with the one calculated by Spitzer Survey of Stellar Structure in Galaxies \citep[S4G,][]{2010PASP..122.1397S}, $-52^{\circ}$, and global properties calculated via their pipeline 3 by \citet{MM15}. 
Note that we can mostly avoid the region with non-circular motion pointed out by \citet{2013ApJ...770L..27F} by using a PVD along the galaxy major axis. We assume that the streaming motion remaining in the PVD is negligible. We can also comment that when considering the error propagation for a simple equation of $v^{2}/2=GM/r$, 10 percent error in the velocity could result in 20 percent error for the SMBH mass, which is consistent the error bar we derive from Figure~\ref{fig:errbars}.

\subsection{Fitting the PVD Model to the Observational Result}\label{subsec:model_fit}
\begin{figure*}
\epsscale{1.0}
\plotone{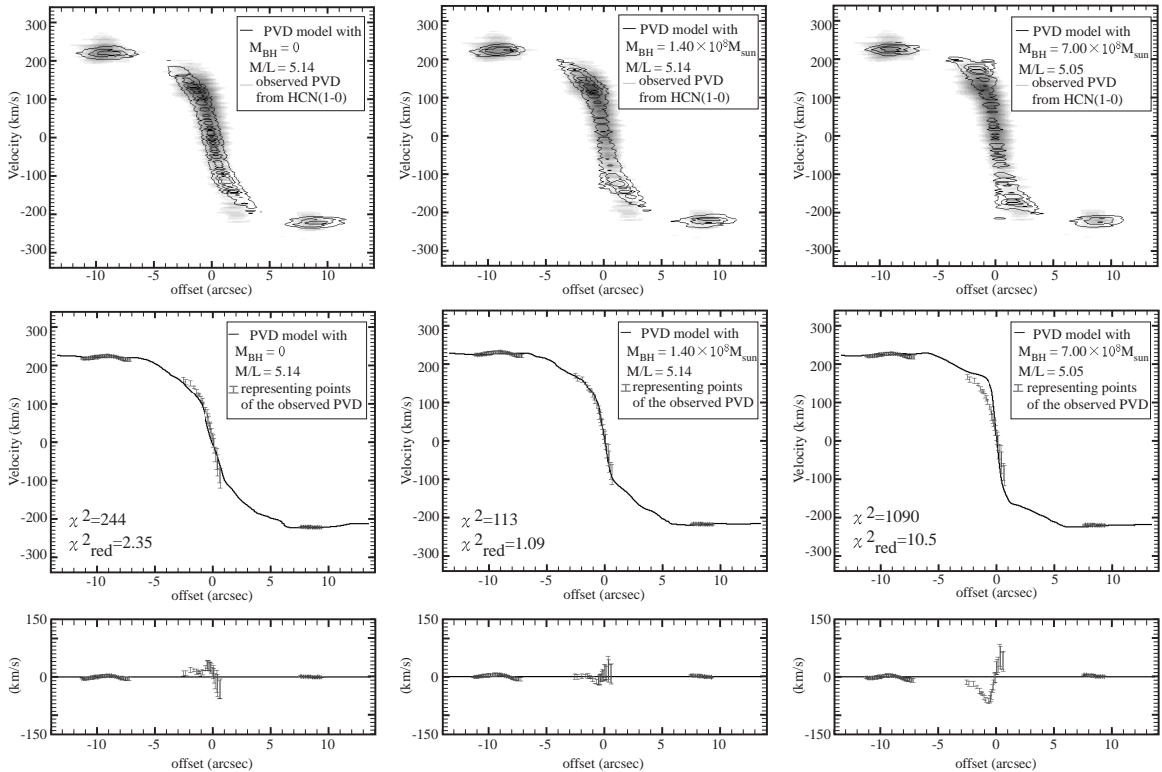}
\caption[Position-Velocity Diagrams and Velocity Fields of Different Parameter Sets]{\small \textit{(Upper panels)} PVDs calculated with the parameter set of $M_{\rm BH}=0$, M/L=5.14 (left), $M_{\rm BH}=1.40\times10^{8}M_{\odot}$, M/L=5.14 (middle), and $M_{\rm BH}=7.00 \times10^{8}M_{\odot}$, M/L=5.05 (right). Black contours are for the modeled PVD, while color filled contours show the observed PVD for ${\rm HCN} (J=1-0)$. 
\textit{(Middle panels)} Each black line shows the PVD model and gray dots show the representative points of the observed PVD. The points are extracted from the gray contours in the upper panel by fitting a Gaussian to the spectrum at each position. Here we fix the lower luminosity limit to represent the observed PVD as $3 \sigma \sim  8.0 {\rm mJy}$. We obtain $\chi^{2}=113$, which is the minimum value, for $M_{\rm BH}=1.40\times10^{8}M_{\odot}$ and M/L=5.14 (middle), while $\chi^{2}=244$ and $1090$ for $M_{\rm BH}=0, {\rm M/L}=5.14$ (left), and $M_{\rm BH}=7.00\times10^{8}M_{\odot}, {\rm M/L}=5.05$ (right). We also put the reduced chi-square value, which is a chi-square value divided with the degree of freedom of 104, as $\chi^{2}_{\rm red}$.
\textit{(Lower panels)} Residual between the black line and plots in the above panels of each. }
\label{fig:PVDfit}
\end{figure*}

The observed PVD of ${\rm HCN} (J=1-0)$ is shown in the upper panel of Figure~\ref{fig:PVDfit} with color filled contours. We set the kinematical position angle to be $130^{\circ} \pm 5^{\circ}$. 
The center of the galaxy is assumed to be the peak of 6~cm observation \citep{1987A&AS...70..517H}, 860 $\mu {\rm m}$, and ${\rm HCN} (J=4-3)$ observation \citep{2013PASJ...65..100I}.
We fit the observed PVD with PVD models calculated with 2 free parameters, the I-band M/L ratio and the SMBH mass. We find the two parameters to be around M/L$\sim$5.0 and $M_{\rm BH} \sim 1.0 \times 10^{8}M_{\odot}$ by initial robust-grid calculation. Then the finer grid of model parameters is set to be from M/L$=4.80$ to 5.35 in steps of 0.01 and $M_{\rm BH}=0.5 \times 10^{8} M_{\odot}$ to $2.5 \times 10^{8} M_{\odot}$ in steps of $0.1 \times 10^{8} M_{\odot}$. 
We calculate an optimal rotation curve from the observed PVD above $3\sigma\sim$8~mJy as follows: we make two cuts -- one in the horizontal and one in the vertical direction. A cut in the vertical direction gives us a spectrum at the pixel whereas a cut in the horizontal direction gives us an intensity profile at that velocity. Peak positions for both are determined with the Gaussian fit. We use the points when the two are consistent, but abandon when they do not match. In the latter case the spectrum is not well characterized by a Gaussian because the asymmetric distribution of the molecular gas and the beam smearing effect is skewing the profile.
106 points are extracted from the observed PVD (see Figure~\ref{fig:PVDfit}). The error bar along the velocity axis for each representative points of the PVD is defined to be a sum in quadrature of the channel width (3.284~km~s$^{-1}$) and the error from Gaussian fitting. After fitting a Gaussian, to the spectrum at each position, we determine the error budget to be the range of all the channels which has an observed value within the difference of half of the noise level from the maximum value of the fitted Gaussian.

\begin{figure}
\epsscale{1.0}
\plotone{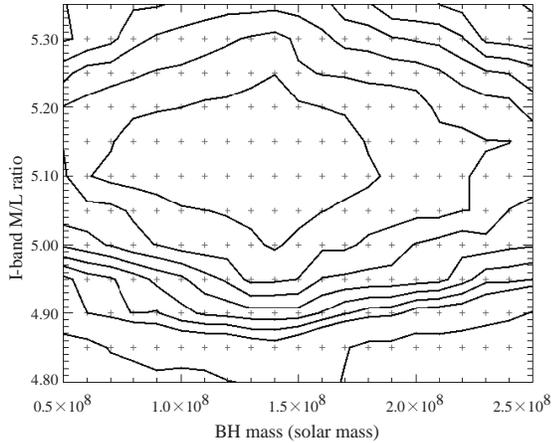}
\caption[Chi-square Distribution for $M_{\rm BH}$ and M/L ratio]{\small Chi-square distribution of the grid calculation with two free parameters, the SMBH mass and the I-band M/L ratio. The inclination angle is fixed to $i=46^{\circ}$ \citep{1989ApJ...342...39O}. Gray dots show the parameter pair for modeling the PVD. The SMBH mass and the M/L ratio are fixed respectively with the range of $M_{\rm BH}=0.50 \times 10^{8} M_{\rm sun}$ to $2.50 \times 10^{8} M_{\rm sun}$ in step of $0.50 \times 10^{8}M_{\odot}$ and M/L=4.80 to 5.35 in step of 0.01. Contour levels are set to be $(2, 3, 4, 5, 6, 8, 12) \times \chi^{2}_{\rm min}$, where $\chi^{2}_{\rm min}=113$ is determined by the minimum chi-square value, realized with parameters of $M_{\rm BH}=1.40 \times 10^{8} M_{\odot}$ and M/L=5.14. We determine the error bar for each parameter by using Figure~\ref{fig:errbars}.}
\label{fig:chisqMLMbh}
\end{figure}

Chi-square values are calculated for each model for the 106 points in the observed PVD. Note that the degree of freedom becomes 104, because we have two free parameters, the SMBH mass and the I-band M/L ratio. Figure~\ref{fig:chisqMLMbh} shows the chi-square value distribution in the parameter space. The contour level is defined to be $(2, 3, 4, 5, 6, 8, 12)\times(\chi^{2}_{\rm min})$.
The smallest chi-square value of 113 (reduced chi-square value is 1.09 when divided with the degree of freedom) is realized with parameters of $M_{\rm BH}=1.40 \times 10^{8} M_{\odot}$ and the I-band~M/L~ratio$=5.14$. See Figure~\ref{fig:PVDfit} to compare three PVD models in black contours and lines calculated with different values of parameters ($M_{\rm BH}=0$, I-band~M/L~ratio$=5.14$ for the left column, $M_{\rm BH}=1.40\times10^{8}M_{\odot}$, I-band~M/L~ratio$=5.14$ for the middle column, $M_{\rm BH}=7.00\times10^{8}M_{\odot}$, I-band~M/L~ratio$=5.05$ for the right column). The red contour in the upper panel shows the observed PVD by ${\rm HCN} (J=1-0)$. Red dots in the middle panel represent the extracted points from the observed PVD. Residuals are plotted in the lower panels. The chi-square values are 244, 113, and 1090 (reduced chi-square values are 2.35, 1.09, and 10.5) for each.

\begin{figure}
\epsscale{1.0}
\plotone{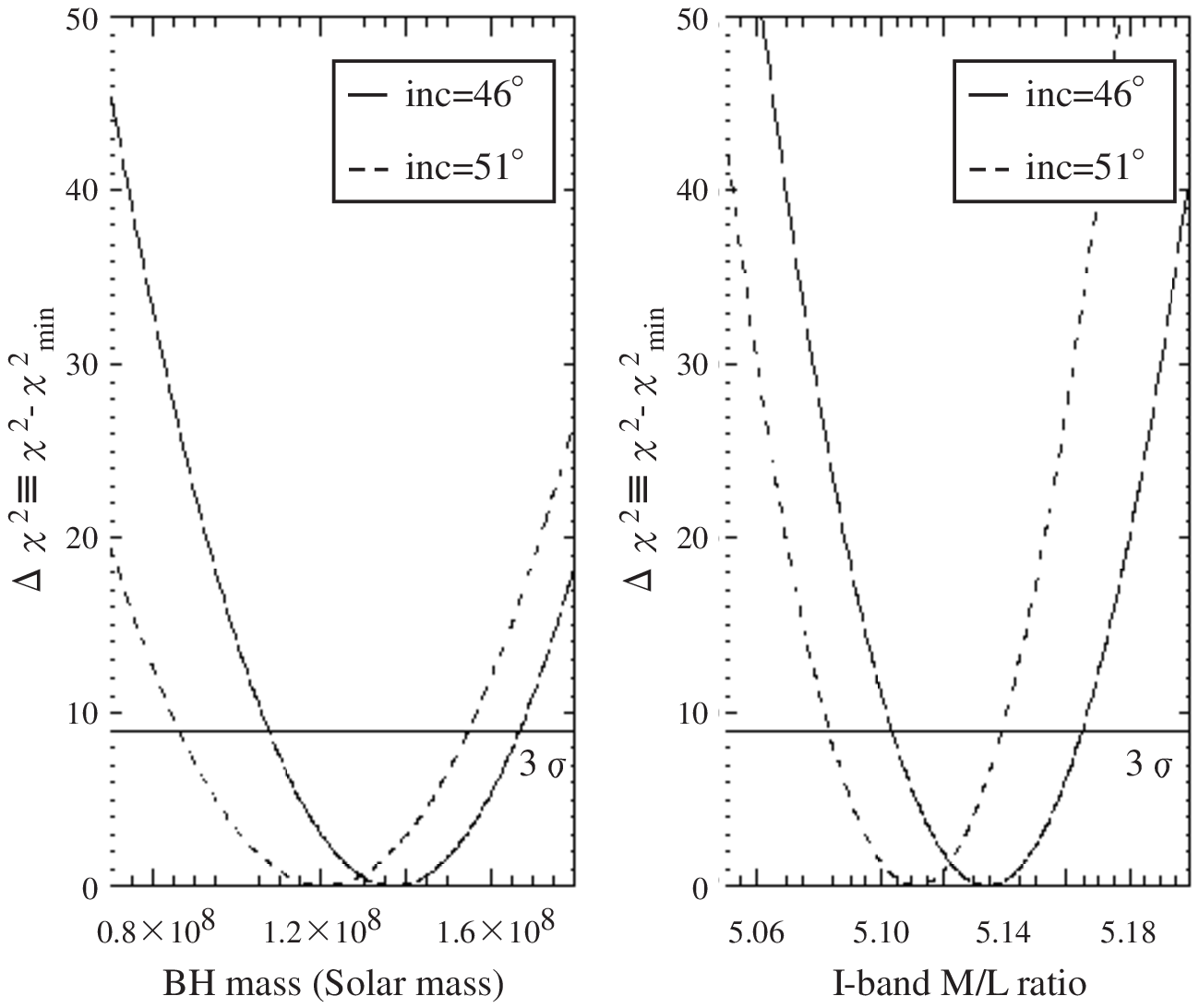}
\caption[Polynomial fittings to the chi-square distribution to determine the error bar]{\small Distributions of the $\Delta\chi^{2} \equiv \chi^{2}-\chi^{2}_{\rm min}$ for each parameter, the SMBH mass (left panel) and the I-band M/L ratio (right panel). Solid black curve shows the distribution calculated at the inclination angle of 46$^{\circ}$. Dashed curve stands for the calculation for 51$^{\circ}$. The error bar at 99.73\% confidence level for each parameter is taken to be the values under solid straight line (3$\sigma$ line, where $\Delta\chi^{2}=9$). For the inclination angle of 46$^{\circ}$, the error bar for each is determined to be $M_{\rm BH}=1.40^{+0.27}_{-0.32}\times10^{8}M_{\odot}$ and M/L$=5.14^{+0.03}_{-0.04}$, while $M_{\rm BH}=1.20^{+0.35}_{-0.34} \times 10^{8} M_{\odot}$ and M/L$=5.11\pm 0.03$ for the inclination angle of 51$^{\circ}$.}
\label{fig:errbars}
\end{figure}

We determine the error bar for each parameter, M/L ratio and SMBH mass, by taking the parameter value within 99.73\% confidence level ($\Delta\chi^{2} \leq 9$, where $\Delta\chi^{2} \equiv \chi^{2}-\chi^{2}_{min}$). Figure~\ref{fig:errbars} show the polynomial fit to the $\Delta\chi^{2}$ distribution for each parameter. In the left panel, the SMBH mass is thus estimated to be $1.40^{+0.27}_{-0.32} \times 10^{8} M_{\odot}$ by considering all the values below the black straight line. The M/L ratio is estimated to be $5.14^{+0.03}_{-0.04}$ by also considering values below the black straight line in the right panel.
Note here that the derived SMBH mass is consistent with the one presumed from the $M_{\rm BH}-\sigma$ relation reported in \citep{2013ApJ...764..184M} and the central velocity dispersion \citep[$196\pm5$~km~s$^{-1}$, ][]{2006ApJ...642..711L}(see also Section~\ref{subsec:intro_n1097}). 

\section{Discussion}\label{sec:dis}
While we determine the SMBH mass and the I-band M/L ratio from the molecular gas dynamics, there are some uncertainties coming from the assumption we made when calculating the model. We discuss the effect of the observing band we use to model the stellar luminosity distribution in Section~\ref{subsec:dis_dust}, how the inclination angle affects the result (Section~\ref{subsec:dis_warp}), and what if we use a different emission line to observe the molecular gas dynamics (Section~\ref{subsec:dis_HCNHCO}).

\subsection{The Proper Stellar Luminosity Profile without the Dust Effect}\label{subsec:dis_dust}
\begin{figure}
\epsscale{0.80}
\plotone{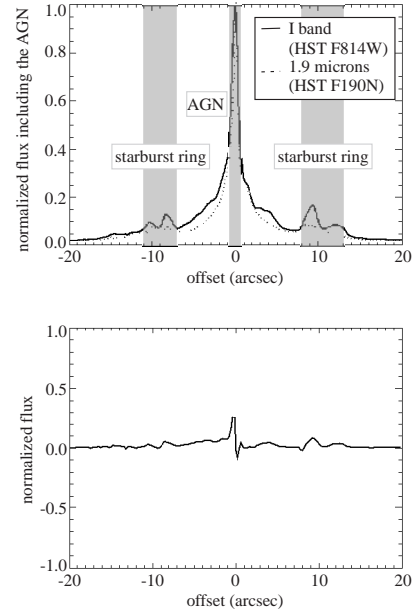}
\caption[\textit{HST} Luminosity Profile for Two Wavelengths]{\small \textit{(Upper panel)} The observed luminosity profile of I-band ( indicated with solid line, \textit{HST} F814W observation) and 1.9 microns (dotted line, \textit{HST} F190N observation). The luminosity value is normalized to compare the intrinsic profile. I-band (F814W) profile is convolved with the assumed point spread function (PSF) of F190N. As discussed in Section~\ref{subsec:model_lumdis}, we shade the starburst ring region and the AGN dominated region.
\textit{(Lower panel)} Residual of the two luminosity distribution showing a good agreement of the two.}
\label{fig:HSTobs_profiles}
\end{figure}

We estimate the expected stellar mass profile by excluding the bright AGN profile and luminosity enhancement by the starburst ring (Section~\ref{subsec:model_lumdis}), but we could not avoid the dust extinction effects, which could be important for this starburst galaxy with its prominent dust lane around the starburst ring. 
One way to mitigate the dust extinction effects is to observe at longer wavelengths such as the near infrared. 
NICMOS on \textit{HST} has a narrow band filter F190N which observes at 1.9 microns. 
We calculate a velocity field from the assumed SMBH mass and the stellar mass profile derived from the luminosity profile of 1.9 microns, and obtain a PVD by following the method described in Section~\ref{subsec:model_PVD}.
We then compare the two PVDs calculated from the F190N luminosity profile and the F814W luminosity profile. 
Both luminosity profiles need to have the same PSF for a proper comparison. 
For the NICMOS data, which have a small field of view and no feasible stars available for a measurement of the PSF, we refer ``Tiny Tim" package \citep[version 6.3][]{2011SPIE.8127E..0JK} and assume the shape as a Gaussian with its FWHM of $0''.4$.
We convolve the I-band luminosity distribution with this PSF. 
 Figure~\ref{fig:HSTobs_profiles} shows the luminosity distribution of the PSF-convolved I-band and 1.9 microns normalized by each maximum value (upper panel) and the difference of two profiles (lower panel). 
This comparison already shows that the difference between the two luminosity profiles is negligible.
We also examine the S4G data at 3.6 microns to compare the stellar profile with the \textit{HST} data but the \textit{Spitzer} Infrared Array Camera (IRAC) PSF is too large ($\sim 1''.8$) to do any detailed comparison.

\begin{figure}
\epsscale{1.0}
\plotone{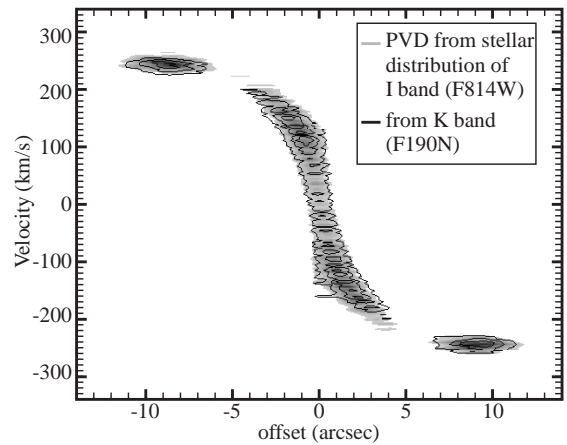}
\caption[PVD by Mass Profiles Derived from Two Luminosity Profiles]{\small Gray filled contours show the PVD calculated from I-band luminosity profile multiplied by the M/L ratio and black contours are the one from 1.9 microns luminosity profile multiplied by the M/L ratio.
Parameters for the calculation is set to be $M_{\rm BH}=1.40\times 10^{8}M_{\odot}$ and M/L=5.14, the best-fit value at the inclination $46^{\circ}$. PVD does not largely differ between the two stellar distribution obtained from different wavelength.}
\label{fig:Kbandvelfield}
\end{figure}

We normalize the peak luminosity at 1.9 microns to the I-band peak, and then subtract the luminosity enhancement of the AGN, as described in Section~\ref{subsec:model_lumdis}.  We also ignore the starburst region and use the same M/L ratio to calculate the velocity field for the two stellar mass profiles.
We find that a $M_{\rm BH}=1.40\times 10^{8}M_{\odot}$ and M/L$=5.14$ gives the best-fit value for an inclination of $46^{\circ}$. The PVDs calculated as such are shown in Figure~\ref{fig:Kbandvelfield}. Black and gray contours in Figure~\ref{fig:Kbandvelfield} are the PVD calculated from two stellar mass profile models -- these do not differ much between the two luminosity profiles. We therefore conclude that the dust extinction effect with the F814W filter is not too serious for the measurement of the SMBH mass.

\subsection{Effect from the Inclination Angle}\label{subsec:dis_warp}
We discuss briefly on the difference coming from how we set the inclination angle. The accuracy of the inclination angle is critical for calculating the velocity and therefore crucial for the SMBH mass estimation. It is however not straightforward to determine the inclination angle when comparing observations at different field of view.
Previous studies of NGC 1097 have determined a dynamical inclination angle of $46 \pm 5^{\circ}$ \citep[HI observations at 3 kiloparsec scales, ][]{1989ApJ...342...39O}, or $34^{\circ}$ \citep[H$\alpha$ line profile study, ][]{2003ApJ...598..956S}.
\citet{2011ApJ...736..129H} reported that the inclination angle of NGC 1097 is $41.7 \pm 0.6^{\circ}$ using the kinematic parameters of $^{12}{\rm CO} (J=2-1)$ observed with Submillimeter Array ({\it SMA}). They argue that the circumnuclear ring is nearly circular for the inclination of $\sim42^{\circ}$, by which means the ring has an intrinsic elliptical shape in the galactic plane, of which case is not symmetric to the axis.
Though the suggested asymmetry is interesting to note, we would like to leave it as a further discussion, since in this work we assume an axisymmetric distribution for stars and molecular gas when calculating the circular velocity field.  
This time we assume the galaxy inclination angle to be $46-51^{\circ}$ by referring to \citet{1989ApJ...342...39O} and the axis ratio of the HST I-band observation. We evaluate the SMBH mass to be $1.40^{+0.27}_{-0.32} \times 10^{8} M_{\odot}$ and the the I-band M/L ratio to be $5.14^{+0.03}_{-0.04}$ at the inclination angle of $46^{\circ}$, with the chi-square value of 113 (1.09 when divided with the degree of freedom 104). 
We also follow the same process in Section~\ref{sec:model} with the inclination angle of $51^{\circ}$ and evaluate the SMBH mass to be $1.20^{+0.35}_{-0.34} \times 10^{8} M_{\odot}$ and the I-band M/L ratio to be $5.11\pm0.03$ with the chi-square value of 117 (1.13 when reduced with the degree of freedom). See also dashed curves in Figure~\ref{fig:errbars} for the chi-square distribution, used to determine the error bar for each parameter.

We can also consider the case of the inclination angle is $34-41^{\circ}$ by multiplying a factor of $\sin(i_{\rm intrinsic})/\sin(46^{\circ}) \sim 0.78-0.91$ to the velocity where we write $i_{\rm intrinsic}$ as an inclination angle of the observed component. Under the simplified assumption that the SMBH mass is proportional to the square of the velocity, we can estimate the change of the SMBH mass to be smaller than $0.31 \times 10^{8}M_{\odot}$, which is mostly included in the error bar of our result $1.40^{+0.27}_{-0.32} \times 10^{8}M_{\odot}$. We therefore consider that it is not crucial to count this error into our error budget.
Note that, however, this galaxy could have a warped or a misaligned structure, which could be interesting to investigate but requires a calculation for an asymmetric potential field.

\subsection{SMBH Mass Estimation from Other Molecular Species}
\label{subsec:dis_HCNHCO}
Our main result is obtained using the HCN line because it had the highest signal-to-noise ratio (SNR). 
It is important to measure the SMBH mass from other molecular species as well for consistency.
We therefore repeat our method using the ${\rm HCO}^{+} (J=1-0)$ emission line. 

\begin{figure}
\epsscale{0.80}
\plotone{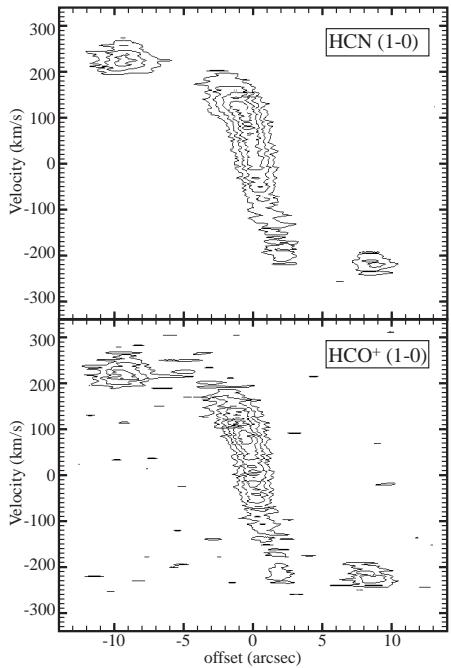}
\caption[Position-Velocity Diagram of ${\rm HCN}(J=1-0)$ and ${\rm HCO^{+}}(J=1-0)$]{\small The observed PVD of ${\rm HCN}(J=1-0)$ (upper panel) and ${\rm HCO^{+}}(J=1-0)$ (lower panel) is respectively shown with black contours. The contour level of both is from $1\sigma$ to $4\sigma$ where  $1\sigma=4.6$~mJy for ${\rm HCN}(J=1-0)$ and $1\sigma=3.2$~mJy for ${\rm HCO^{+}}(J=1-0)$. The velocity structure of these two PVDs are in good agreement.}
\label{fig:PVDHCNHCO}
\end{figure}

We apply the fitting procedure described in Section~\ref{subsec:model_fit} to the PVD for ${\rm HCO}^{+}(J=1-0)$, and estimate the SMBH mass to be $(1.40 \pm 0.30) \times 10^{8} M_{\odot}$ and the I-band M/L ratio to be $5.15 \pm 0.03$ with a galaxy inclination of $46^{\circ}$. These derived values are consistent with the measurement using ${\rm HCN} (J=1-0)$. From Figure~\ref{fig:PVDHCNHCO}, we see that the observed PVDs of two molecular gases are in good agreement, indicating that the fitting parameters will be consistent between the two.

Reaching the velocity structure from multiple molecular species is one of the particular benefit of the SMBH mass measurement with millimeter/submillimeter wavelength observation, which enable one to observe more than two molecular species at the same frequency band.

\section{Conclusion}\label{sec:conc}
We derive the SMBH mass in NGC 1097 to be $1.40^{+0.27}_{-0.32} \times 10^{8}M_{\odot}$ by using dense molecular gas dynamics traced with ${\rm HCN} (J=1-0)$ and  ${\rm HCO^{+}}(J=1-0)$ observed with ALMA.
The value of SMBH mass is measured with two emission lines is in good agreement, indicating the applicability of this method to any nearby galaxy with detectable molecular gas.
Furthermore, the mass is consistent with $M_{\rm BH}-\sigma$ relation \citep[][]{2013ApJ...764..184M} from the velocity dispersion observed by \citet{2006ApJ...642..711L}. We can comment that the derived mass does not coincide with the  $M_{\rm BH}-\sigma$ relation for early-type galaxies, but for mixed samples and for late-type galaxies.
This time we consider that the dust extinction effect is not very crucial to model the luminosity distribution of the galaxy, but the inclination angle could affect the SMBH mass when we observe the rotational motion with better resolution.
As millimeter/submillimeter interferometers develop their angular resolution and sensitivity, this method will provide more samples, especially late-type galaxies with their central gas dynamically well relaxed, to correlations between SMBH mass and galaxy properties such as $M_{\rm BH}-\sigma$ relation.
Increasing the number of galaxy samples in  $M_{\rm BH}-\sigma$ relation will lead us to higher accuracy of the correlation, which suggests the coevolution process of galaxies and BHs. 

\acknowledgments
The authors would like to express gratitude to Dr. S. Komugi and Dr. Y. Matsuda, who provided constructive comments and suggestions concerning this paper. The authors would like to acknowledge Prof. M. Bureau and Dr. M. Cappellari for fruitful discussions. This paper makes use of the following ALMA data: ADS/JAO.ALMA\#2011.0.00108.S. ALMA is a partnership of ESO, NSF (USA), and NINS (Japan), together with NRC (Canada) and NSC and ASIAA (Taiwan), in cooperation with the Republic of Chile. The Joint ALMA Observatory is operated by ESO, AUI/NRAO, and NAOJ. The NRAO is a facility of the National Science Foundation operated under cooperative agreement by Associated Universities, Inc.
This work is also based on observations made with the NASA/ESA \textit{HST}, obtained from the Data Archive at the Space Telescope Science Institute, which is operated by the Association of Universities for Research in Astronomy, Inc., under NASA contract NAS 5-26555.
Data analysis were in part carried out on common use data analysis computer system at the Astronomy Data Center, ADC, of the National Astronomical Observatory of Japan.
A part of this study was financially supported by JSPS KAKENHI Grant Number 26*368.

\clearpage
\end{document}